# Predicting Star Scientists in the Field of Artificial Intelligence: A Machine Learning Approach

Koosha Shirouyeh[1], Andrea Schiffauerova[2], and Ashkan Ebadi[2,3,*]

[1] Mechanical, Industrial, and Aerospace Engineering Department, Concordia University, Montreal, QC H3G 2W1 Canada
[2] Concordia Institute for Information Systems Engineering, Concordia University, Montreal, QC H3G 2W1 Canada
[3] National Research Council Canada, Toronto, ON M5T 3J1, Canada

[*] Email: ashkan.ebadi@nrc-cnrc.gc.ca

**Abstract** Star scientists are highly influential researchers who have made significant contributions to their field, gained widespread recognition, and often attracted substantial research funding. They are critical for the advancement of science and innovation, and they have a significant influence on the transfer of knowledge and technology to industry. Identifying potential star scientists before their performance becomes outstanding is important for recruitment, collaboration, networking, or research funding decisions. Using machine learning techniques, this study proposes a model to predict star scientists in the field of artificial intelligence while highlighting features related to their success. Our results confirm that rising stars follow different patterns compared to their non-rising stars counterparts in almost all the early-career features. We also found that certain features such as gender and ethnic diversity play important roles in scientific collaboration and that they can significantly impact an author's career development and success. The most important features in predicting star scientists in the field of artificial intelligence were the number of articles, group discipline diversity, and weighted degree centrality. The proposed approach offers valuable insights for researchers, practitioners, and funding agencies interested in identifying and supporting talented researchers.

## 1. Introduction

Star scientists are individuals who have made exceptional contributions to their respective fields, characterized by high publication and citation rates, groundbreaking research contributions, and influential collaborations (Hirsch, 2005; Ioannidis et al., 2014). Identifying and understanding the characteristics of star scientists is crucial for advancing research and applications in various fields, as well as recognizing and supporting exceptional researchers (Azoulay et al., 2011). Star scientists are vital as they drive innovation, push the boundaries of knowledge, and often serve as catalysts for breakthrough advancements. They often have the expertise and influence to bring together researchers from different fields and foster interdisciplinary collaborations, leading to new insights and breakthroughs in research (Lee and Bozeman, 2005; Uzzi et al., 2013).

Artificial intelligence (AI) is a rapidly evolving field characterized by constant innovation and transformative technologies, making it dynamic and ever-changing. AI has a wide range of applications across industries, from healthcare (Mosallaie et al., 2021; Song et al., 2023) to finance (Cao, 2022) and transportation (Abduljabbar et al., 2019; Iyer, 2021), impacting society at large. Considering AI's complex and constantly evolving nature, it necessitates a broad spectrum of expertise from various scientific domains. Therefore, the effective dissemination of knowledge can promote the exchange of insights and skills among AI scientists, aiding them in overcoming prevalent challenges when tackling intricate real-world problems (Hajibabaei et al., 2023).

Predicting star scientists in the field of AI is important since it allows for early identification and support of exceptionally talented individuals, aiding in recruitment, collaboration, and research funding decisions. This can lead to the acceleration of groundbreaking research and innovation in the field of artificial intelligence, contributing to its continued growth and impact on society. Motivated by the importance of early recognition of talents and the special characteristics of AI as a rapidly evolving field, in this work, we propose a machine learning model that accurately predicts the individuals destined to become star scientists in the field of AI. The model leverages a comprehensive set of features of different types, including network measures, diversity metrics, and research output, to enhance the accuracy of the prediction task. Additionally, the study seeks to discern the key differentiators between rising star scientists and their peers by comparing early-career characteristics, thereby contributing to a deeper understanding of the factors that drive success in AI research.

The majority of existing approaches for identifying rising stars primarily target individual researchers. Nonetheless, scientific research frequently thrives on collaborative endeavors, and an individual's success can be intertwined with the success of their collaborations. Therefore, there is a need to explore methodologies that not only assess individual researchers but also take into account their collaborative efforts and the dynamic interactions within their scientific networks. This work makes significant contributions to the field by addressing the critical issue of identifying star scientists in the AI domain using machine learning techniques. The study not only offers a predictive model for recognizing these influential researchers at an early career stage but also sheds light on the unique patterns and features that distinguish rising stars from their non-rising counterparts, providing valuable insights into the characteristics and patterns of successful researchers. Furthermore, the research highlights the impact of various factors, including gender and ethnic diversity, on scientific collaboration and an author's career trajectory, underscoring their significance in the development and success of researchers. Notably, the study identifies the key features that are most instrumental in predicting star scientists. These findings provide valuable insights for researchers, practitioners, and funding agencies seeking to proactively identify and support promising talents, ultimately advancing the field of artificial intelligence and fostering innovation and knowledge transfer to industry.

The paper structure is as follows. It begins by reviewing the related works outlining the significance of predicting star scientists. The "Data and Methodology" section describes the data, predictive model and its feature set. The "Results" section presents the findings, emphasizing the distinctions between rising star scientists and their counterparts and identifying key factors contributing to their success. The paper concludes by discussing the broader implications of these findings for researchers, practitioners, and funding agencies interested in nurturing talent in the field.

## 2. Related Work

## 2.1.    Definition of Star Scientists

The literature presents a wide range of definitions for star scientists, highlighting the diverse nature of scientific excellence. These definitions encompass various criteria, such as citations, funding, patenting activity, and membership in prestigious scientific organizations, reflecting the multifaceted ways in which researchers achieve star status.  Lowe and Gonzalez-Brambila (2007) considered stars as those highly productive scholars that become entrepreneurs. In another study, faculty members who founded new technology ventures were considered star scientists (Hess and Rothaermel, 2012). Patents have been also used to identify star researchers. In a study conducted by Niosi and Queenton (2010), researchers holding over five patents and producing at least one major publication per year were counted as stars. Schiffauerova and Beaudry (2011) focused on the field of biotechnology and identified stars as those with at least 20 patents. Moretti and Wilson (2014) also focused on patents and identified stars as patent assignees whose patent count over the past decade ranks within the top 5% of patent assignees on a national scale. In a more comprehensive definition, Azoulay et al. (2010) defined star researchers as those satisfying at least one of the following criteria: 1) researchers with substantial funding, 2) extensively cited scientists, 3) leading patent holders, and 4) individuals elected as the National Academy of Science members. Hess and Rothaermel (2011) used a quantitative criteria and defined star scientists as researchers whose publication and citation rates exceeded the mean by three standard deviaitons. Another set of studies established star scientists by considering citation metrics. For example, Hoser (2013) considered academics with the highest citation counts. Or, Tartari et al. (2014) defined stars as academics within the top 1% for citations in their field and within the top 25% for grants received from Engineering and Physical Sciences Research Council (EPSRC) in the United Kingdom. In a more recent study in Japan, star scientists were characterized as researchers featured in the Highly Cited Researchers (HCR) published by Clarivate Analytics (Nagane et al., 2018).

## 2.2.    Importance of Star Scientists

Star scientists are crucial players to the advancement of research and applications in various fields (Wagner and Leydesdorff, 2005). They can drive innovation and interdisciplinary collaborations, often leading to new insights and breakthroughs in research (Wuchty et al., 2007). Recognizing and supporting exceptional researchers is therefore critical for the progress of science and technology (Azoulay et al., 2019). As decribed in the previous section, star scientists are often characterized by high publication and citation rates, groundbreaking research contributions, and influential collaborations (Lee and Bozeman, 2005). They assume leadership roles in research domains, offering direction and inspiration to their peers. Moreover, they are often the recipients of prestigious awards and honors, which can have a significant impact on their professional journeys and research trajectories (Bornmann, 2014).

The impact of star scientists extends beyond their individual achievements. Research indicates that they make disproportionate contributions across diverse contexts (O'Boyle Jr. and Aguinis, 2012) and engage in collaborations with a broder spectrum of scientists (Abramo et al., 2019). They could also play a gatekeeper role, facilitating the exchange of knowledge within several research groups and impacting neighboring researchers in terms of productivity and recognition (Azoulay et al., 2010; Oettl, 2012). Stars not only affect academia but also have a significant influence on businesses, transfering advanced knowledge to emerging technology firms through different channels such as co-founding or advisors roles (Zucker et al., 1998). Their work frequently leads to the development of innovative technologies and applications, carrying substantial economic and societal benefits. In addition, they actively engage in policy formulation and public outreach, contributing to the formation of public discourse and opinion on important matters (Leshner, 2003). These all highlights the vital role of star scientists in the scientific ecosystem and society.

## 2.3.    Predicting Star Scientists

Throughout their career, researchers may transition through various phases such as declining, emerging as a rising star, maintaining stability, or achieving well-established status, based on their performance (Tsatsaronis et al., 2011). Certain researchers maintain consistent success throughout their professional journeys while others exhibit fluctuating patterns. Productive scientists often receive increased recognition, which, in turn, fuels their future productivity. Consiquently, star scientists are expected to have a dominant profile in the early stages of their career, capitalizing on the principle of accumulative advantage to ascend to star status (Azoulay et al., 2010). Given the importance of star scientists, predicting rising stars in academia has emerged as an active area of research in recent years, leading to the development of various approaches and methodologies.

Utilizing bibliographic data, various techniques such as social network analysis (SNA) and machine learning (ML) have been employed for the identification of rising stars. Initially, researchers endeavored to use the dynamics of author ranking based on compiling a list of each author's important scores to perform social network analysis to find star scientists (Daud et al., 2013; Li et al., 2009). There have also been studies that used ML techniques to identify potential future star scientists. For instance, Daud et al. (2015) built ML classifiers, considering increases in citation counts along with three kinds of characteristics, i.e., author, venue, and co-authorship, to forecast future rising stars, underscoring the significance of the publication venues in the prediction process. Regression models have also been deployed, accounting for both temporal and content variables (Zhang et al., 2017). Their research unveiled that temporal factors, as opposed to venue features, serve as the most influential predictors of rising stars. These studies rely on citation counts as the assessment metric for identifying rising stars, which may not provide a reliable signal of a researcher's overall success (Nie et al., 2019). Nie et al. (2019) proposed a new approach to overcome this limitation. They considered multiple factors, including the quantity and impact of articles, citation counts, the domain cited factor, and the co-impact authors to generate a composite score for each scientist. Next, they utilized changes in this composite score across two consecutive five-year intervals to label rising stars. Their methodology demonstrated superior performance compared to prior studies relying on citation growth. Additionally, their research highlighted the importance of the venue characteristic in the identification of rising stars.

Despite the numerous approaches proposed for identifying rising stars in the literature, there are still unaddressed research gaps. One notable gap is the absence of a universally reliable and comprehensive method for pinpointing star scientists. A universal scientific performance evaluation indicator is foundational to crafting an all-encompassing approach for identifying rising stars. While Nie et al. (2019) introduced a composite score that draws from various indicators to overcome the limitations of single-aspect indicators, e.g., citation counts, they acknowledged that their approach requires a wide range of metadata encompassing citations, coauthors, and venues which  can add complexity to the evaluation of scientific performance. In Addition, several studies have examined the correlation between current research impact and different attributes. For example, using SNA, significant positive associations were found between degree centrality and the h-index (Abbasi and Altmann, 2011). Furthermore, studies have explored diversity metrics, underscoring a robust connection between diversity and scientific performance in general (AlShebli et al., 2018). In another line of research, previous studies (e.g., Daud et al., 2015; Nie et al., 2019) have compared the predictive outcomes between different categories of features, identifying the significance and predictive power of those features. However, a more holistic approach could involve the consideration of feature combinations from diverse categories, as opposed to evaluating each category in isolation.

In summary, identifying rising stars in science is a complex task that requires considering a wide range of factors. Although existing approaches have made significant progress, several research areas remain unaddressed, and there is a need for more accurate predictive models that can handle large-scale data and incorporate diverse features. This study aims to address some of these gaps by developing a machine learning prediction model, trained on large-scale data, to predict star scientists using a combination of features of different types. We also investigate early-career features and compare rising and non-rising star scientists, examining how current characteristics of academics can influence their future success.

## 3. Data and Methodology

Figure 1 illustrates the high-level conceptual flow of the analyses. AI-related publications were collected from the Scopus database, while data about publishers and journals were sourced from Scimago. After merging the data, it underwent several pre-processing steps. Relevant metadata, including research themes, author gender and ethnicity, network measures, and diversities, were extracted. Rising star and non-rising star junior AI scientists were identified based on their scientific impact over two consecutive periods. Several machine learning classifiers were trained and tested. Feature importance was examined to propose a combination of features from different categories crucial for predicting future star scientists. The pipeline was coded in Python programming language. Detailed explanations of all the modules will be provided in the following sections.

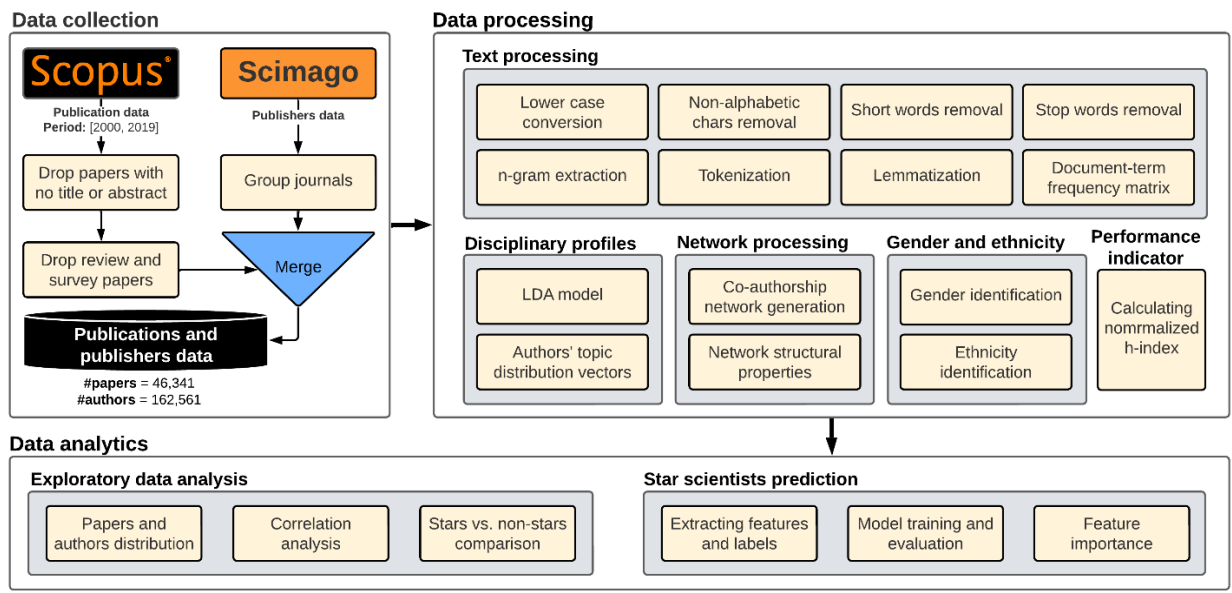

**Figure 1.** The high-level conceptual flow of the analyses.

## 3.1. Data

In this research, we utilized the same dataset that was used in our previous work (Hajibabaei et al., 2022). The dataset encompassed articles related to artificial intelligence (AI) published between 2000 and 2019, sourced from Elsevier's Scopus database. The extraction process was facilitated using the following query: ("artificial intelligence" OR "machine learning" OR "deep learning"). The data was filtered to include research articles, conference papers, book chapters, and books. We opted for Scopus as our data source of choice, given its remarkable retrieval rate of computer science articles and its extensive indexing of unique content within the field (Cavacini, 2015). Furthermore, we enriched the database by incorporating journal ranking data obtained from SCImago. We further incorporated historical publisher metadata, sourced from the SCImago, to gauge publications, classifying them according to their respective publisher's SCImago Journal Rank (SJR) at the point of publication. The SJR indicator serves as a metric to assess the academic eminence of scholarly journals, factoring in the volume of citations received by a journal as well as the prestige of the journals from which these citations originate. Publications were stratified into three tiers contingent on the SJR of their respective publishers at the time of publication. Journals with an SJR exceeding three standard deviations above the mean were categorized as "A", those with an SJR surpassing one standard deviation but falling short of three standard deviations above the mean were designated as "B", and journals with an SJR below one standard deviation above the mean were denoted as "C". The dataset contains 46,341 publications and 162,561 authors.

## 3.2. Methodology

### 3.2.1. Textual data processing

The paper's title and abstract serve as concise representations of key research elements, encapsulating specialized keywords and core research concepts in a publication (Ebadi et al., 2020). As the first step, we merged the titles and abstracts of the publications in the dataset and conducted a series of preprocessing steps on the combined textual content. These steps encompassed converting text to lowercase, filtering out short words (fewer than 3 characters), removing custom stop words, eliminating non-alphabetic characters, lemmatization, and tokenization. Unigrams and bigrams were also extracted and a document-term frequency matrix was generated. The processed data was employed later to extract researchers' disciplinary profiles, as described in Section 3.2.2.

### 3.2.2. Feature engineering

Exploring individual attributes of authors is essential for unraveling patterns within the academic landscape. Notably, gender disparities persist in academia, an issue that extends to the community of star scientists as well (Sá et al., 2020). Additionally, it is evident that star scientists exhibit a heightened propensity for international collaborations in contrast to their non-star counterparts (Abramo et al., 2019). Moreover, various diversity metrics, including ethnic diversity, exhibit significant associations with scholars' academic accomplishments (AlShebli et al., 2018). Evaluating these factors requires extracting additional information not present in the original dataset. We created engineered various features of different types to enhance the performance of the machine learning models, allowing for a more comprehensive data representation. Different The feature engineering process and the feature set are described in detail in this section.

### 3.2.2.1. Gender

Gender of the authors in the dataset were identified and added to the original data using the approach proposed in our previous work (Hajibabaei et al., 2022), where we combined machine learning and natural language processing techniques to build an automated gender assignment model. This model underwent training on an extensive dataset annotated with gender labels, enabling the inference of gender based on a set of core features encompassing full names, affiliations, and countries of origin.

### 3.2.2.2. Ethnicity

We used the "ethnicolr" packge in Python programming language to determine ethnicity of the authors in the dataset. The package employs machine learning techniques to categorize any given name into one of the 13 ethnic groups listed below:

- Asian, Greater East-Asian, East-Asian
- Asian, Greater East-Asian, Japanese
- Asian, Indian Sub-Continent
- Greater African, Africans
- Greater African, Muslim
- Greater European, British
- Greater European, East-European
- Greater European, Jewish
- Greater European, West-European, French
- Greater European, West-European, Germanic
- Greater European, West-European, Hispanic
- Greater European, West-European, Italian
- Greater European, West-European, Nordic

### 3.2.2.3. Diversity

Diversity and its implications have been extensively studied (e.g., Hong and Page, 2004; Woolley et al., 2010). Academic collaborations, which often involve researchers from varied geographical locations, fields of expertise, and backgrounds, provide insights into the structure of scholarly cooperation (Deville et al., 2014; Jia et al., 2017). The growing interest in analyzing collaboration networks has driven efforts to understand the underlying factors influencing academic achievement (Fortunato et al., 2018). Several of these factors, such as academic discipline (Hajibabaei et al., 2023), gender (Hajibabaei et al., 2022), career age (Jones and Weinberg, 2011), ethnicity (Freeman and Huang, 2015), and institutional affiliation (Jones et al., 2008), have been investigated, and their impact on research performance has been documented. In this work, we investigated five classes of diversity as follows:

- **Ethinicity diveristy:** It takes the ethnic backgrounds of researchers into the account.
- **Age diversity:** The term age pertains to the academic age of researchers in the dataset, calculated based on the date of their first publication. We considered the following four sub-categories for this feature:
  - **Group 1:** 0-5 years of experience.
  - **Group 2:** 5-10 years of experience.

         ◦ **Group 3:** 10-15 years of experience.
         ◦ **Group 4:** 15-20+ years of experience.
- **Gender diversity:** This category considers the gender of researchers in the collaboration networks.
- **Disiplinary diversity:** It was measured both at individual and group levels. At the group level, it involved examining the diversity of co-authors' areas of expertise, which corresponded to the most probable topics in each author's disciplinary profile. At the individual level, this diversity was defined by considering the variety of fields represented in an author's publications.
- **Affiliation diversity:** This category accounts for the countries listed in the affiliations of co-authors in a paper and aims to assess the extent of international collaboration diversity. We used Shannon entropy (Shannon, 1948) as the diversity metric.

### 3.2.2.4. Disciplinary profiles

To extract researchers'disciplinary profiles, we applied a topic modeling technique to the document-term frequency matrix, outlined in Section 3.2.1. Specifically, we built a Latent Dirichlet Allocation (LDA) topic model (Blei, 2003) to extract the main topics in the publications. To find the optimal number of topics, we built a series of LDA baseline models, each employing varying number of topics. These models were subsequently evaluated using metrics such as perplexity and log-likelihood (Griffiths and Steyvers, 2004). Inter-topic distance mappings were also visually assessed. In addition to quantitative metrics, a qualitative evaluation was conducted by examining keywords and document-topic distributions within the models. Through a combined quantitative and qualitative analysis, the optimal number of topics was determined to be 8. Subsequently, leveraging the LDA model, each publication was associated with multiple topics. The authors' research domains were established based on the average topic distribution across their publications, facilitated by the document-topic probability matrix. Each author was finally represented by an topic distribution vector of length 8, with each component reflecting the average thematic distribution of the author's papers within that topic. These disciplinary profiles were added to the original dataset.

### 3.2.2.5. Network structural features

Collaboration is a fundamental aspect of scientific activities, serving as a conduit for researchers to share their expertise, resources, and insights to address intricate challenges and achieve significant breakthroughs (Sonnenwald, 2007). The position of researchers within collaborative networks can influence their overall performance (Ebadi and Schiffauerova, 2015a, 2016a). Although quantifying and representing scientific collaboration is a complex task, co-authorship is the widely accepted standard for measuring collaboration (Ebadi and Schiffauerova, 2015b; Price, 1963; Ubfal and Maffioli, 2011), as a robust indicator of mutual scientific engagement. In co-authorship networks, each node represents an individual researcher, while the presence of a link between two nodes indicates that these two researchers have collaborated on at least one shared publication. We employed Pajek software (De Nooy et al., 2018) to create the co-authorship networks of researchers across each year within the study period. Subsequently, we calculated the following network structural variables at the individual level of researchers:

- **Betweeness centrality:** Betweenness centrality serves as a metric for evaluating a researcher's potential influence on network communication (Abbasi and Altmann, 2011). Researchers possessing high betweenness centrality can act as conduits between disparate research communities, regulate the dissemination of information, and wield more influence over their peers within the network. This influence extends to activities such as determining project priorities and facilitating the spread of knowledge (Ebadi and Schiffauerova, 2016b). This measure is computed by dividing the total number of shortest paths by the fraction of shortest paths passing through a particular node among all node pairs (Borgatti, 2005), and ranges between 0 and 1. The most central nodes, often called as *gatekeepers*, exhibit the highest betweenness centrality values.

- **Degree centrality:** Degree of a node is calculated based on the count of connections the node has (Diestel, 2016). Respectively, the degree centrality of a node is established according to the degree of the node, and the resulting values are normalized to fall within the range of 0 to 1. Within co-authorship networks, researchers with high degree centrality are indicative of higher activity, as they possess a greater number of direct connections (Wasserman and Faust, 1994). Having a larger number of direct collaborators could facilitate the researcher's access to diverse sources of skills and complementary expertise (Ebadi and Schiffauerova, 2016b).

- **Weighted degree:** The link weight between two nodes signifies the strength of their collaborative connection, representing the frequency of their collaborations. We calculated this measure for each researcher as the sum of the weights associated with a node's direct connections, divided by the total number of distinct co-authors. Researchers with strong ties, often collaborating with the same partner, are considered to be loyal connections (Abbasi and Altmann, 2011).

- **Clustering coefficient:** The clustering coefficient assesses the inclination of nodes to assemble into clusters and quantifies the number of triangles, thereby indicating the extent of clustering (Ebadi and Schiffauerova, 2016b). Essentially, it gauges the probability that two neighbors of a node are interconnected with each other (Carrington, 2011). Researchers exhibiting a high clustering coefficient create closely-knit clusters, potentially enhancing their capacity to generate superior research by leveraging the strong interconnections within their groups and facilitating internal referring among team members (Ebadi and Schiffauerova, 2016b). The clustering coefficient is calculated by dividing the actual number of edges between neighbors by the maximum possible number for the network under consideration.

### 3.2.2.6. Research performance

Various metrics are commonly used to assess scientists' productivity and the impact of their publications. These metrics encompass indicators such as impact factor, total document count, citation count, citations per document, and the number of highly cited publications. The h-index, introduced by Hirsch in 2005, combines measures of a researcher's publication quantity and impact into a single metric (Hirsch, 2005). The h-index has found utility across a spectrum of fields, including biology (Bornmann and Daniel, 2005), information science (Cronin and Meho, 2006), and journal evaluation (Braun et al., 2006), to name a few. The key strength of the h-index lies in its ability to amalgamate measures of quantity and impact into a comprehensible and applicable metric, hence; it retains its practicality as a proxy for evaluating authors' research performance within domains focused on authors from the same discipline. In this work, we selected the h-index as an indicator to gauge the research performance of authors at different stages of their careers.

### 3.2.3. Data analytics

#### 3.2.3.1. Feature set and label

As described in Section 3.2.2., we generated several features of different types and added them to the original dataset. Features that were used in the analyses are as follows: 1) Number of publications, 2) Number of publications based on journal ranking, 3) Citation count, 4) h-index, 5) Individual disciplinary diversity, 6) Group disciplinary diversity, 7) Ethnic diversity, 8) Gender diversity, 9) Affiliation diversity, 10) Age diversity, 11) Degree centrality, 12) Wighted degree centrality, 13) Clustering coefficient, and 14) Betweenness centrality.

To investigate the progression of junior researchers into star scientists, this research focuses on authors who published their first publication between 2006 and 2010, marking the initial decade of their professional journeys, and had at least one collaboration (n=9,391 authors). The selection of a 10-year timeframe aligns with common practice in the analysis of rising stars (e.g., Daud et al., 2015; Nie et al., 2019), facilitating a more comprehensive exploration of their developmental trajectory during the early-career phase. By concentrating on researchers who commenced their careers within a similar time frame, this approach enables a meaningful comparison of their relative success while considering potential confounding variables like historical context and technological advancements that could have influenced the career trajectories of authors who began their professional journeys at different periods. To generate the label, we considered rising stars as researchers in the dataset whose h-index experienced a notably substantial increase, i.e., exceeding the average by at least three standard deviations. This growth rate is calculated by contrasting an author's h-index during the initial and subsequent five-year periods of their career, as stated in Equation (1):

$$H_{GR} = \frac{(h_2 - h_1)}{(t_2 - t_1)} \qquad (1)$$

where $h_1$ and $h_2$ represent the h-index during the first five years and first ten years, and $t_1$ and $t_2$ denote the time of the first and second periods, respectively. Having generated the label, we built machine learning models to predict rising stars and analyze their early-career characteristics such as research productivity, diversity indicators, and network structural features during the first five years of their career.

#### 3.2.3.2. Classification

We defined the star scientist prediction problem as a supervised learning task in which the machine learning model learn from labeled training data. More specifically, the problem was defined as a binary classification task, where the label was either 0 (non-rising) or 1 (rising star). The dataset consists of authors who had their first publication between 2006 and 2009 (n=7,311) for the training set and authors with their first paper in 2010 (n=2,313) for the unseen test set. To address the imbalanceness in the distribution of rising stars and non-rising stars within the training set, we utilized the Synthetic Minority Over-sampling Technique (SMOTE) (Chawla et al., 2002). SMOTE aids in rebalancing the classification by augmenting the number of samples in the minority class while preserving the inherent data distribution characteristics.

To ensure models'generalizability and reduce the risk of overfitting, we used an expanding window cross-validation approach for hyperparameter tuning and training the models. The expanding window technique derives its name from the fact that the training set's size increases progressively as the validation set shifts forward in time, initiating with a small validation set and gradually enlarging it, providing the model with a progressively larger dataset for learning. In particular, we implemented an expanding window cross-validation, wherein the data was divided into five equally sized (1-year) windows of data, with the fifth window was set aside as the unseen test set. The validation procedure commenced with the first window as the training set and the second window as the validation set. Subsequently, the training set was progressively extended to encompass the second window (i.e., validation set in previous iteration), and the third window was taken as the validation set. This process was repeated until the entire dataset was employed for validation. This approach allowed us to evaluate the performance of the model over time, ensuring that the generalizability and high performance of the models.

In the feature selection process, we employed Recursive Feature Elimination (RFE) to identify the most pertinent features. The objective of RFE is to reduce data dimensionality by iteratively excluding the least significant features while preserving the most critical ones. RFE proves particularly advantageous when handling high-dimensional datasets, enhancing classifier performance by eliminating extraneous or redundant features and guarding against overfitting, a situation in which a classifier becomes overly intricate and underperforms on new, unseen data. We applied RFE to each of the classifiers to enhance their performance and mitigate the risk of overfitting. We trained and built four classification models, under the defined validation and feature selection settings, as follows: 1) Logistic Regression (LR) (Hosmer Jr et al., 2013), 2) Support Vector Machine (SVM) (Cortes & Vapnik, 1995), 3) Gaussian Naive Bayes (NB) (Zhang, 2004), and 4) Random Forest (RF) (Breiman, 2001). The performance of classifiers were evaluated using F1 score and area under the ROC curve (AUC). The F1 score, a metric that strikes a balance between precision and recall, offers a holistic measure of classifier performance, taking into account the count of both false positives and false negatives in predictions (González, 2010). The AUC score assesses the model's capacity to differentiate between positive and negative classes at various thresholds, considering both true positive and false positive rates. It offers a unified measure that gauges the overall predictive quality of the model, providing a more comprehensive evaluation of its performance, which was particularly valuable in our case due to the imbalanced nature of the data.

# 4. Results

## 4.1. Distribution of publications and authors

Figure 2 shows the distribution of papers, authors, and the number of authors per paper in the dataset. As seen in Figures 2-a and 2-b, the data reveals a remarkable growth in both the number of publications and the number of authors in the field of AI from 2000 to 2019. Specifically, the number of papers has surged from 91 in 2000 to over 16,000 in 2019, indicating an exponential increase in research output. Correspondingly, the number of authors has also expanded dramatically, from 194 authors in 2000 to nearly 54,000 authors in 2019. This trend may reflect the burgeoning interest and investment in AI research, as well as the increasing collaboration and interdisciplinary efforts within the scientific community. The sharp rise in publications and authors suggests a vibrant and rapidly evolving field, with a growing pool of researchers contributing to the advancement of AI. This expansion underscores the importance of developing robust methods to identify and support rising star scientists amidst the increasing volume of research activity.

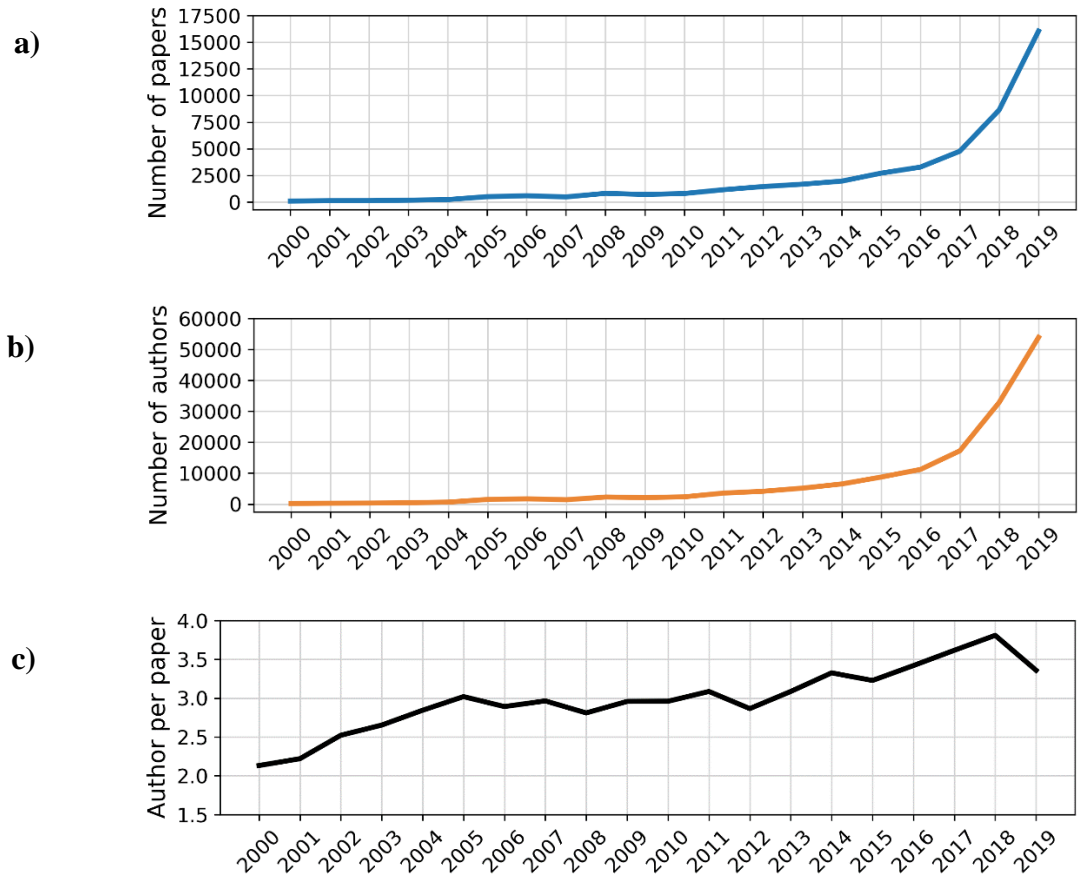

**Figure 2. a)** Distribution of publications, **b)** distribution of authors, and **c)** average number of authors per paper.

The observation of the average number of authors per paper increasing from around 2 in 2000 to approximately 4 in the later years (Figure 1-c) further highlights the growing trend of collaborative research in the field of artificial intelligence. As seen in the Figure, from 2000 to 2005, there was a noticeable rise in the average number of authors per paper, followed by a steady trend until 2012, and then another phase of increasing collaboration. The initial rise and subsequent steady trend might reflect the early stages of increased interdisciplinary collaboration and team-based research becoming more common in AI. The renewed upward trend post-2012 may suggest a further intensification of collaborative efforts, possibly driven by the increasing complexity of research problems that require diverse expertise and larger research teams. This shift towards more collaborative research practices underscores the importance of fostering teamwork and interdisciplinary approaches in scientific endeavors, which can be crucial for making significant advancements in AI.

### 4.2. Correlation analysis

Before predicting star scientists, we investigated the correlations between the feature set and the label, i.e., the growth rate of h-index. As illustrated in Figure 3, a positive correlation exists between the h-index growth rate and the number of articles. Additionally, there is a positive correlation between the h-index, number of articles, weighted degree centrality, and group discipline diversity within the first five years of a researcher's career. Moreover, correlations were observed between diversity indicators, performance, and network structural metrics. For instance, gender diversity demonstrated positive associations with degree centrality, weighted degree centrality, and clustering coefficient, with the highest correlation seen between ethnic diversity and clustering coefficient in these two feature groups. This may imply that individuals from diverse ethnic backgrounds tend to form close-knit relationships and establish dense connections, resulting in a high clustering coefficient early in authors' careers. Similarly, it may suggest that authors collaborating with individuals of various genders, as opposed to exclusively those of the same gender, tend to have larger networks of connections, underscoring the significance of gender diversity in scientific collaboration and its potential impact on authors' career growth. Lastly, as anticipated, the number of level-A articles was positively correlated with citation count, indicating that publishing in prestigious venues garners more recognition, which can result in a higher number of citations.

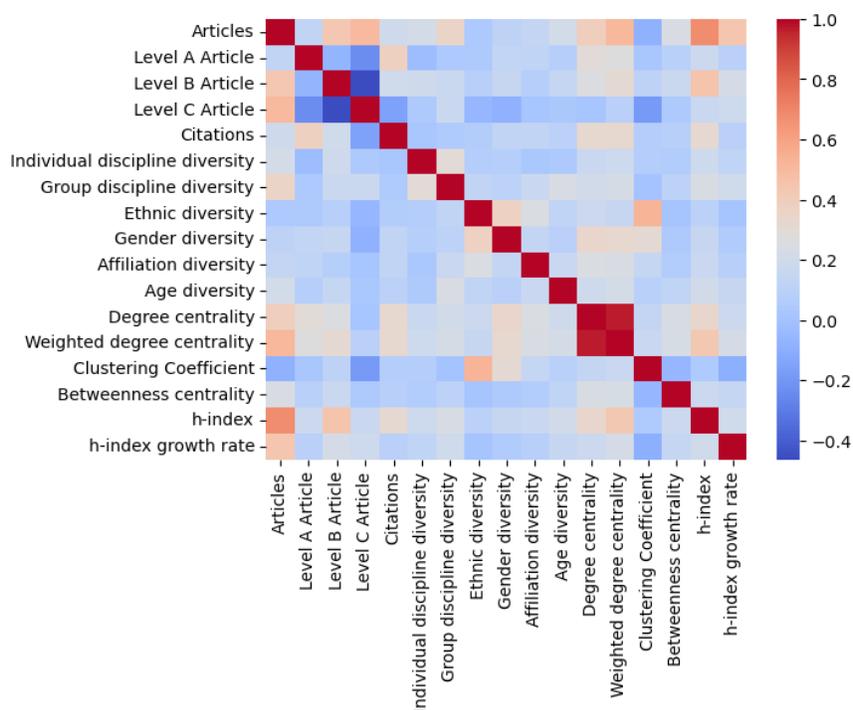

**Figure 3.** Correlation heatmap.

Conducting a pairwise comparison hypothesis testing, we further investigated whether rising stars possessed different early-career characteristics compared to the general population of scientists. By implementing a two-sample t-test, we tested the null hypothesis that there were no significant differences in the means of the two groups across various features. Our analysis revealed significant differences in early-career characteristics between rising star scientists and the general population of scientists, with the exception of ethnic diversity. This may indicate that rising stars follow unique trajectories in their early careers, which can be predictive of their future success. This finding may further underscore the importance of recognizing and nurturing specific early-career attributes that contribute to scientific excellence while also highlighting that ethnic diversity remains consistent across different levels of scientific achievement.

### 4.3. Star vs. non-star researchers

We performed a comparative analysis of the features between rising stars and non-rising researchers in the dataset to gain initial insights into the factors influencing their success. As seen in Table 1, rising stars exhibit a higher average weighted degree compared to their average degree, whereas non-rising stars demonstrate nearly equivalent average values. Moreover, rising stars exhibit higher averages in weighted and unweighted degrees in comparison to non-rising stars. This may suggest that rising stars engage more and frequently in repeated collaborations with the same researchers, as indicated by higher weighted degree values. This may reflect the formation of strong and productive collaborations, potentially playing a pivotal role in their success (Wuchty et al., 2007). Additionally, rising stars exhibit a tendency to publish more frequently and attract citations from a broader spectrum of publications in their early careers when contrasted with non-rising stars. This implies their enhanced ability to garner recognition for their contributions, with a higher average number of high-impact articles published. These initial comparisons underscore the significance of collaboration and the impact of publications in shaping a prosperous research career. Balancing the production of substantial work with a focus on its quality, recognition from diverse sources, and effective, productive collaborations emerges as crucial.

**Table 1.** Rising star vs. non-rising star researchers.

| Feature | Rising stars (label=1) | | | | Non-rising researchers (label=0) | | | |
|---|---|---|---|---|---|---|---|---|
| | **Mode** | **Mean** | **σ** | **Range** | **Mode** | **Mean** | **σ** | **Range** |
| Articles | {1,3} | 2.82 | 1.74 | [1,10] | 1 | 1.25 | 0.67 | [1,13] |
| Level-A articles | 0 | 0.29 | 0.6 | [0,3] | 0 | 0.1 | 0.32 | [0,4] |
| Level-B articles | 0 | 1.35 | 1.34 | [0,6] | 0 | 0.45 | 0.67 | [0,12] |
| Level-C articles | 1 | 1.18 | 1.44 | [0,9] | 1 | 0.7 | 0.74 | [0,12] |
| Citations | 10 | 42.89 | 41.27 | [0,237] | 0 | 19.63 | 35.1 | [0,757] |
| Individual discipline diversity | 0.2 | 0.19 | 0.03 | [0.09,0.25] | 0.18 | 0.16 | 0.03 | [0.04,0.25] |
| Group discipline diversity | 0 | 0.15 | 0.14 | [0,0.36] | 0 | 0.04 | 0.1 | [0,0.37] |
| Ethnic diversity | 0.35 | 0.25 | 0.1 | [0,0.37] | 0 | 0.23 | 0.14 | [0,0.37] |
| Gender diversity | 0 | 0.23 | 0.12 | [0,0.37] | 0 | 0.17 | 0.15 | [0,0.37] |
| Affiliation diversity | 0 | 0.15 | 0.13 | [0,0.37] | 0 | 0.09 | 0.13 | [0,0.37] |
| Age diversity | 0 | 0.12 | 0.12 | [0,0.35] | 0 | 0.05 | 0.1 | [0,0.37] |
| Degree | {3,4} | 10.8 | 9.56 | [1,63] | 0 | 5.2 | 4.89 | [0,57] |
| Weighted degree | 4 | 13.12 | 12.02 | [1,76] | 3 | 5.58 | 5.59 | [0,68] |
| Clustering Coefficient | 0.07 | 0.05 | 0.03 | [0,0.15] | 0.07 | 0.06 | 0.03 | [0,0.26] |
| Betweenness centrality | 0 | 0.0002 | 0.0006 | [0,0.005] | 0 | 0.0002 | 0.0002 | [0,0.001] |
| h-index | 1 | 1.79 | 0.95 | [0,5] | 1 | 0.58 | 0.58 | [0,8] |

| h-index growth rate | 0.6 | 0.74 | 0.24 | [0.6,2.2] | 0 | 0.04 | 0.09 | [0,0.4] |
|---|---|---|---|---|---|---|---|---|

### 4.4. Predicting Star Scientists

As detailed in Section 3.2.3.2., we built four classifiers to predict rising star scientists and evaluated their performance using F1 and AUC metrics. As shown in Figure 4, the random forest classifier outperformed the others, achieving an AUC of 0.75. While the performance difference between the random forest and the SVM model was notable, it was not overwhelmingly substantial. However, a key advantage of the random forest model over SVM is its reduced susceptibility to overfitting, especially when confronted with high-dimensional datasets. The random forest model constructs numerous decision trees on random subsets of the training data and averages their outcomes, which helps to mitigate overfitting and enhances the model's generalization to unseen data. For classification, the features selected using the RFE method included the number of articles, citation count, individual discipline diversity, ethnic diversity, gender diversity, weighted degree centrality, clustering coefficient, and betweenness centrality. These features provided a foundation for distinguishing rising star scientists from their peers.

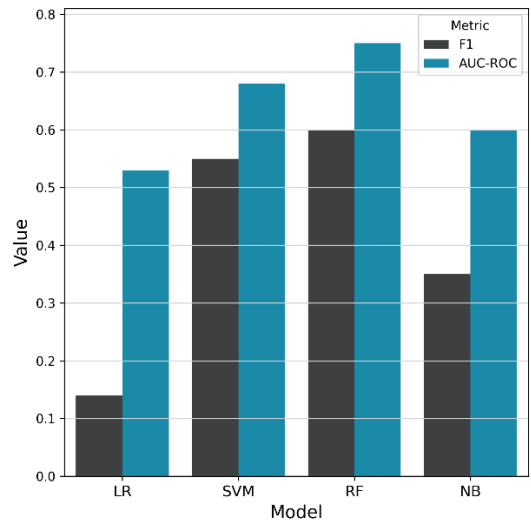

**Figure 4.** Performance comparison of classifiers.

## 5. Conclusion

Artificial intelligence is gaining significant attention and interest worldwide due to its unique characteristics and capabilities. It offers the promise of automating and enhancing tasks that were once considered exclusive to human intelligence, spanning from natural language processing and computer vision to problem-solving and decision-making. The availability of massive digital data and increased computational power have driven advancements in AI algorithms, leading to more accurate and complex models. Consequently, AI technologies are being applied across various domains such as healthcare (Ebadi et al., 2021; Ma et al., 2023), finance (Cao, 2022), and even entertainment (Hallur et al., 2021), making it a versatile and interdisciplinary field. The ability to create systems that continuously learn and improve from data, known as machine learning, is a core aspect of AI, offering adaptability and scalability in addressing diverse challenges. This potential is attracting immense interest and investment in the field.

Predicting rising star scientists is crucial for several reasons. Firstly, identifying rising stars can help research institutions and funding agencies allocate their resources more efficiently by investing in individuals who are likely to make significant contributions to the field. Secondly, these predictions benefit junior researchers themselves by offering insights into the factors that drive success, allowing them to tailor their career paths. Additionally, the scientific community at large benefits because star scientists often lead groundbreaking research, contributing to technological advancements with profound societal impacts. Similarly, predicting AI star scientists is crucial because it allows organizations to identify and support emerging talent, fostering innovation and ensuring a competitive edge in the rapidly evolving AI landscape. Additionally, it helps allocate resources effectively, driving breakthroughs and advancements in AI research and applications.

This research extends the literature on the utilization of machine learning algorithms for forecasting the trajectories of early-career scientists, underscoring the potential of such techniques in advancing our comprehension of the intricate scientific landscape. By analyzing features of various types, we developed a predictive model that distinguishes rising stars from their non-rising counterparts. We focused on the field of artificial intelligence and harnessed diverse datasets, encompassing publication and citation data, and applied various techniques, including NLP and social network analysis, to engineer various features of different types such as gender, field of expertise, ethnicity, and network structural measures. It was found that rising stars exhibit notable distinctions across various attributes in comparison to the broader research population. The pivotal features for effective predictions encompassed article quantity, citation count, individual discipline diversity, ethnic diversity, gender diversity, weighted degree centrality, clustering coefficient, and betweenness centrality. Our findings also indicated that certain metrics, such as the number of published articles, group discipline diversity, and weighted degree centrality, are significant indicators of future success. These insights may provide a foundation for more informed recruitment, collaboration, and funding decisions, which are critical for fostering innovation and scientific advancement. Additionally, our research highlighted the importance of diversity in the scientific community. Gender and ethnic diversity were found to play crucial roles in collaboration dynamics and career development. The inclusion of diverse perspectives not only enriches the research environment but also enhances the potential for groundbreaking discoveries. This underscores the need for policies and initiatives that promote diversity and inclusion within scientific institutions and funding bodies, thereby ensuring a more equitable and productive scientific landscape.

Overall, the proposed approach can serve as a valuable tool for researchers, practitioners, and funding agencies aiming to identify and support emerging talent in AI. By recognizing and nurturing star scientists early in their careers, the scientific community can accelerate the transfer of knowledge and technology to industry, driving progress and innovation. Future research should continue to refine predictive models and explore additional features that contribute to a researcher's success, further enhancing our ability to support and advance the frontiers of science.

## 6. Limitation and Future Work
Several constraints should be acknowledged when interpreting the findings of this study. The dataset utilized in this research was restricted in both the timeframe and available metadata, which could impact the precision of the research performance metric due to the absence of comprehensive citation data. Subsequent investigations might extend the data collection period and incorporate supplementary metadata from diverse sources. Additionally, researchers have the opportunity to enhance the research performance metric by integrating alternative indicators to offer a more comprehensive evaluation of the researcher's work's influence and dissemination.

Another potential avenue for future research involves exploring the influence of mentorship and collaboration on the success of junior researchers. While this study considered collaboration diversity in the success of early-career researchers, there remains an opportunity to delve deeper into the specific forms of collaborations and mentorship that contribute to academic success. This could entail conducting surveys or interviews with accomplished scholars and their mentors to gain insights into the types of collaborations and mentorship that prove most advantageous. Furthermore, forthcoming research might assess the impact of the academic environment on the accomplishments of early-career scientists, encompassing factors like institutional resources available to researchers and the prevailing research community culture.

Lastly, it could be of great interest to replicate this research in various academic disciplines, aiming to explore the determinants of research achievement within diverse fields of study. Such an approach may unveil shared elements and distinctions in the factors influencing research success, thereby offering valuable insights into the most effective strategies for nurturing and promoting the growth of rising stars across academia.